# Distributional Ground Truth

Non-Redundant Crowdsourcing Data Quality Control in UI Labeling Tasks

Maxim Bakaev[*]

Novosibirsk State Technical University, Russia

Sebastian Heil[*]

Technische Universität Chemnitz, Germany

Martin Gaedke

Technische Universität Chemnitz, Germany

HCI increasingly employs Machine Learning and Image Recognition, in particular for visual analysis of user interfaces (UIs). A popular way for obtaining human-labeled training data is *Crowdsourcing*, typically using the quality control methods *ground truth* and *majority consensus*, which necessitate redundancy in the outcome. In our paper we propose a non-redundant method for prediction of crowdworkers' output quality in web UI labeling tasks, based on homogeneity of distributions assessed with two-sample Kolmogorov-Smirnov test. Using a dataset of about 500 screenshots with over 74,000 UI elements located and classified by 11 trusted labelers and 298 Amazon Mechanical Turk crowdworkers, we demonstrate the advantage of our approach over the baseline model based on mean Time-on-Task. Exploring different dataset partitions, we show that with the trusted set size of 17-27% UIs our "distributional ground truth" model can achieve $R^2$s of over 0.8 and help to obviate the ancillary work effort and expenses.

## 1 INTRODUCTION

Many human-computer systems developed today incorporate AI technologies, which increasingly rely on machine learning (ML) datasets [1]. The threshold of data volume that allows AI to become effective for most practical tasks is estimated as millions of labeled data samples [2]. This is not always attainable in HCI, but for instance user behavior models that are gaining in popularity in the field can be fairly hungry for training data. They say that data are big and ubiquitous nowadays, but relevant and high-quality data is not so easy or cheap to obtain, particularly if it involves human activities [3]. Overall, the principal sources of training data are:

---

[*] Both authors contributed equally to the work

- Open-sourced datasets. Currently, they are mostly tailored to a specific task of a particular research group and thus unlikely to match requirements of a concrete user interface (UI) design.
- Automated collection from the web. Even if relevant data can be found, usually humans need to structure, correct and filter it, which still involves considerable effort.
- Artificial data. Augmenting the data allows extending data volumes at low costs, but it so far has been bound only to certain types of tasks (e.g. image recognition [4]).
- Annotated/labeled data. In most domains, appropriately trained and motivated humans can produce the best training data, but this process is generally time-consuming.

Understandably, relatively few research groups and development teams can afford full-time annotators, so data labeling tasks are often outsourced. This rationalizes the emergence and rapid development of crowdsourcing (crowdworking) platforms lately: Amazon Mechanical Turk (AMT) (2005), microworkers.com (2009), Yandex.Toloka (2014), Google's AutoML (2018), etc. Generally, ML-related Human Intelligence Tasks (HITs) requested through the platforms involve unskilled and tedious work in data gathering and processing, so the crowdworkers' wages and motivation remain rather low [5]. Correspondingly, the core challenge in crowdsourcing today is obtaining data of appropriate quality [6], and the platforms struggle to support data quality assessment and control [7]. This is accompanied by the emergence of related meta-tools, such as CDAS, Crowd Truth, iCrowd, DOCS for AMT, and so on [8]. Another trend is the platforms that specialize in tasks of a particular kind or from a specific domain, e.g. Mighty AI, Hive (.ai), Scale (.ai), etc. for AI in automotive industry [9]. In our study we address UI labeling task that is gaining in popularity as computer vision methods are seeing wider application in HCI, particularly for UI visual analysis [10], but for which no dedicated quality control methods had been developed, to the best of our knowledge.

Of the crowd data quality control methods, majority/group consensus (MC) and ground truth (GT) are arguably the most widely used, being also supported in most of the platforms [7]. These two methods necessarily imply redundancy (several workers performing the same task), which remains the mainstream approach in the crowdsourcing quality assessment [11]. Besides wasting some (up to 67% for the MC) of the potentially useful work effort, redundancy is problematic for certain types of tasks, particularly the ones implying unstructured responses or no strictly right or wrong answers. Development of non-redundant data quality control methods has the potential to decrease the share of low-quality or unnecessary data that essentially goes to waste.

Such methods can involve comparing the results or some auxiliary parameters to common sense or to a "truth" outside of the task at hand. An example of such auxiliary parameter is Time-on-Task, one of the popular factors in crowdsourcing quality control [7]. Obviously, knowing some trusted characteristics that the new data should comply with in order to be of desired quality is potentially even more advantageous. The idea of the current paper was inspired by Zipf's law ability to distinguish between natural language texts, in which the words' frequencies largely comply with the Zipf's (power law) distribution [12], and random texts that seemingly do not exhibit this phenomenon [13]. Since user interface is a message that a human and a computer exchange in their interaction, we might expect it to have similarly good fit to Zipf's/power law, or at least the frequencies of UI elements to be equally characteristical of its significance.

So, in our paper we propose and test the "distributional ground truth" (DGT) method – an approach for non-redundant crowdsourcing data quality control. The rest of the paper is organized as follows. In Section 2, we provide overview of the related work and the involved apparatus, including Kolmogorov-Smirnov test for 2 samples, which is the statistical foundation of the DGT method. In Section 3, we describe our experimental study which involved three sessions: 1) with the 11 trusted labelers, 2) with the 22 verifiers of the trusted labelers' output, and 3) with 298 AMT crowdworkers. In Section 4 we analyze the data, apply the DGT method and compare the results to the baseline and the alternative factors in data quality control. In the final Section, we discuss the results, note the limitations of our findings and outline directions for further research.

## 2 METHODS AND RELATED WORK

### 2.1 Crowdsourcing in HCI and the UI Labeling Task

Various taxonomies of crowdsourcing tasks have been developed lately, e.g. in [14], [15], [16], etc. In [17], they identified 8 dimensions of crowdsourcing in software engineering that allow defining specific crowdsourcing models such as peer production, competition and microtasking. Crowdsourcing for software creation based on a platform for crowd-supported creation of composite web applications combining passive and active crowdsourcing has been proposed in [18]. In the HCI field, crowdsourcing has been applied e.g. for solving small UI design problems at a large leveraging diversity of microtasking results in CrowdDesign [19], to adapt web layouts to a variety of different screen sizes in CrowdAdapt [20], and so on.

Figure 1 presents spider diagram of the 8 dimensions for the general microtasking and for the UI labeling (that show high overlap), the values for the latter being obtained with the argumentative approach as described in [17]. One of the arguable dimensions is Expertise Demands, since labeling of UIs, unlike general images, requires certain investment in learning about the task. On the other hand, the expertise is far from being comparable to a professional one, e.g. a doctor marking signs of tumor in X-Ray images. So, as we estimate the pre-investment into UI labeling expertise as 15-20 minutes on average needed to read and comprehend the provided instructions, we decided to assign the value of 1 to this scale.

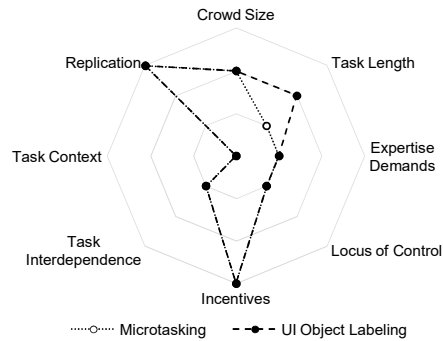

Figure 1: The eight dimensions of general microtasking and UI labeling

Choosing proper HIT design is essential for obtaining high-quality results from tasks that involve human factors [15]. The two principal designs possible for UI labeling HITs are "locate all instances of UI elements of a particular class" or "locate all UI elements and specify class for each one". The former would be more familiar to crowdworkers, since it resembles general image labeling, and probably be faster overall. The latter appears more appropriate to UI labeling for HCI ML purposes (particularly due to importance of nested UI elements) and can allow more efficient data quality control.

## 2.2 Crowdsourcing Data Quality Control

A recent and comprehensive review of quality control methods for crowdsourcing is provided in [7], where the methods are organized into three major groups: individual, group and computation-based. The former two generally imply involvement of humans into assessment of the annotators or of the tasks output, thus suggesting additional overhead in the work effort. Among the computation-based methods, which can be performed by machines, of particular interest with respect to our study are the ones that rely on statistics.

A trendy direction of research within this sub-field is supplementing the two most popular quality control methods for more effective aggregation of the results or allocation of the workers. For instance, in [11] the authors proposed statistical quality estimation based on a two-stage probabilistic generative model for crowdsourcing tasks implying unstructured output. In [21] they studied optimal distribution of training set answers, and it was shown that accuracy of majority voting is highest if the labels in training data follow a uniform distribution. An extension of ground truth method based on probability distributions was proposed in [4], where the decision whether to run a second crowd labeling on an image depends upon the "trusted" distribution of labels.

The two relevant general statistical approaches are outlier analysis and interlaboratory comparison [22]. An example of the former applied to crowdsourcing is [23], where they evaluated performance of UIs. The authors did not find statistically significant differences for the data collected in the lab settings and through AMT. This however required resolving certain practical challenges, particularly in outlier detection. The methodology of the *ISO 13528:2015 Statistical methods for use in proficiency testing by interlaboratory comparison* standard is mostly built upon the "allowed error", but it also includes comparison of the measurements to some known distributions [22, ch. 8.4].

In the distributional analysis for quality assessment of data related to human activities, the use of Benford's law in fraud detection arguably remains the most prominent example. The approach was extended with certain degree of success to power laws in general, suggesting their application in biometrics, forensics and network traffic analysis [24]. Still, comparing the assessed samples to Zipf's law, which is the foremost example of a power law, is most convincing in text analysis. Considerable amount of work in the field base quality control on goodness of fit to Zipf's distribution: for peer reviews [25], text coherence [13], etc. Unfortunately, comparison to power law requires relatively high number of values in the distribution, which can be hard to attain in crowdsourced HITs. Particularly, in the considered UI labeling task that would imply the number of distinct labeling classes to greatly exceed the number of commonly used UI elements' types. The requirement towards the distribution size can be to some extent mitigated through simulation and reliance on the goodness-of-fit (GOF) measure instead of the p-value that would be invariantly low. For the GOF testing we are going to rely on the method described in [26] and the corresponding software library *plpva.r*, which actually also employs KS test (but for a single sample) with simulation.

## 2.3 The "Distributional Ground Truth" (DGT) Method

Whether a phenomenon shall exhibit truly good fit to a particular distribution is unpredictable: e.g., power laws are not that common outside of natural language analysis. Instead, we believe that a method that does not assume fit to a known distribution, but instead relies on empirical distribution function acting as the ground truth, could see wider applicability.

With respect to the classifications provided in [7], the method that we propose falls into such categories as "data quality", "accuracy", "computation-based" and "outlier analysis". The principal idea is testing of distributions' equality (homogeneity), i.e. calculation of statistical distance between the trusted labelers' results and the assessed sample (not exactly a *sample*, as the numbers are not sampled from a population). In this, we do not mean the *distributions* as probability distributions or generalized functions, but rather as frequency distributions. By design, the method is appropriate for classification with a limited number of classes, whose frequencies are used as the values of the variable and need to vary considerably. The same requirement implies that the classified material is reasonably consistent and systematic, so that it could be largely described with a rationed "vocabulary" of classes. Hence, the pre-requisites of the method appear valid for the UI labeling task that we consider in our work, although it would require enough results for each worker for composing a representative distribution.

The distance measure appropriate for the purposes of the method can be provided by Kolmogorov-Smirnov (KS) nonparametric test for two samples. The test compares the samples' cumulative distributions and computes p-value that depends on the largest discrepancy (distance) between the distributions [27]. Being more powerful than Mann-Whitney's test used to compare the medians of two unpaired groups of data, KS test is sensitive to differences in both **location** and **shape** of the distributions. The null hypothesis is that both samples are randomly drawn from the same set of values. Among the assumptions of KS test for two samples are:

- the samples are mutually independent,
- the scale of measurement is at least ordinal,
- the variables are continuous.

Of the two statistics provided by KS test, we are going to rely on p-values, since the distance measures D may have different degrees of freedom and are not directly comparable. Since the values for the KS distribution functions are known and tabulated, the computational complexity of the test is defined by the sorting stage, and thus is not worse than $O(n^2)$, some algorithms even reducing it to $O(n)$ [28]. Another advantage of the KS test is that it can work with relatively low number of values in the two distributions, unlike for testing the fit to power law. Actually, when the samples sizes are close, like we plan to have, increasing one sample may lead to the paradoxical higher bias in the KS test [29].

## 3 THE EXPERIMENTAL STUDY DESCRIPTION

The objective of the study is to explore the effectiveness of the method in crowdsourced data quality control and to estimate the efficient size of the trusted set. The hypothesis is that the DGT method can be used to better explain performance of crowdworkers in UI labeling tasks compared to the baseline or the alternative factors we are about to consider. An important note is that although the trusted labelers and the crowdsourcers had worked with the same material, the study design ensures that the trusted and the testing sets never overlap, so redundancy does not emerge.

### 3.1 The Trusted Set Labeling

The objective of our first experimental session was to obtain the trusted set that could provide the "distributional ground truth" in our study.

#### 3.1.1 Material

The material for the UI labeling was screenshots of homepages of websites belonging to higher educational organizations (universities, colleges, etc.). Initially, we collected 10639 screenshots in PNG format using the dedicated Python script crawling through URLs that we acquired from various catalogues (DBPedia, etc.). To ensure better diversity of UI elements, the screenshots were made for full web pages, not just of the part above the fold or of a fixed size. Then we manually selected 497 screenshots for the experiment using the following criteria:

- University or college corporate website with reasonably robust functionality;
- Not overly famous university;
- Website content in English and reasonably diverse (i.e. no photos-only websites);
- Reasonable diversity in website designs (colors, page layouts, etc.).

*3.1.2 Participants*

The trusted labelers were student members of the Novosibirsk State Technical University crowd-intelligence lab, who volunteered to work in the project and provided informed consent. In total, there were 11 of them (6 male, 5 female), with age ranging from 20 to 24 (mean = 20.5, SD = 0.74). All the labelers had normal or corrected to normal vision and reasonable experience with web UIs and IT.

*3.1.3 Procedure*

To perform the task, the participants used LabelImg (https://github.com/tzutalin/labelImg), a third-party dedicated software tool they were asked to install on their computers. It allows drawing bounding rectangle around an image element, specifying a label for it, and saving the results as XML files. The screenshots were distributed among the participants near evenly, but no random assignment was performed. The labelers worked independently and on their own computer equipment, and each of them was provided with the identical instruction containing the following main points:

1. After loading the next UI screenshot, click *Create \nRectBox* and locate as many UI elements as you can, by bounding them with the rectangles. Do not cut the labeled elements' borders, but also try not to grab any empty space, except when the element is not rectangular. The rectangles can be adjusted or duplicated, if necessary.
2. Classify each of the UI elements by choosing its type from the list of pre-defined classes (see in Table 1). If none of the pre-defined classes seem appropriate, you can add a custom class.
3. After all the visible UI elements are located and classified, save the results (PASCAL VOC format) and move to the next assigned UI screenshot.

*3.1.4 Design*

We have devised the list of 20 labeling classes for the UI elements. In that, we sought to cover the three major groups of visual objects specific for web UIs: graphical content elements, textual content elements, and interface elements. The names and descriptions of the classes are presented in Table 1, and were provided to the labelers with the instruction.

The output of the labeling was the collection of XML files, each corresponding to its UI. As per Pascal VOC format, for each labeled UI element there was the specification of the bounding box *(xmin-top left, ymin-top left, xmax-bottom right, ymax-bottom right)* and the name of the class. Using dedicated scripts, we derived the following variables for each of the 11 labelers:

- distribution of classes, i.e. the number of labels in each class (both pre-defined and custom);
- mean number of labeled elements per UI: $EUI_T$.

As the labeling was performed at the participants' convenience, we did not measure the Time-on-Task.

## 3.2 The Labeling Quality Verification

The objective of the second experimental session was to obtain the assessments of the trusted labelers' performance.

### 3.2.1 Material

The verification was performed for the UIs produced by the trusted labelers. Each labeled UI was represented as the combination of the screenshot file (exactly the same as the trusted labelers used) and the Pascal VOC XML file containing the labeling results. These were rendered together in dedicated web-based software, which would add the verification information to the XML.

Table 1: UI elements classes used by the trusted labelers

| # | Class Label | Description |
| --- | --- | --- |
| 1 | **image** | foreground images that the web page displays |
| 2 | **backgroundimage** | images that are used as background, i.e. other UI Elements are placed on top of them and they have no semantic meaning |
| 3 | **panel** | an area that is visually separated from its surroundings by borders, shadows, and/or background color and contains at least one other UI element |
| 4 | **list** | any list (numbered or unnumbered) that uses bullet points, numberings, borders, background color etc. to display a set of similar items |
| 5 | **table** | any visually recognizable table (using alignment, lines or background color to represent rows and columns) |
| 6 | **paragraph** | a portion of text consisting of one or more lines of text that are not visually separated by white space and/or indentation from other text |
| 7 | **textblock** | two or more subsequent paragraphs of text |
| 8 | **text** | any other portion of text that is neither a label nor a paragraph or textblock |
| 9 | **symbol** | any graphical symbol, can appear on buttons, tabs, links, in texts etc. or separately |
| 10 | **checkbox** | must be labeled one-by-one, without the accompanying text (which must be marked as label) |
| 11 | **radiobutton** | must be labeled one-by-one, without the accompanying text (which must be marked as label) |
| 12 | **selectbox** | a listbox that would expend when clicked, displaying several options which can be selected or multi-selected |
| 13 | **textinput** | single line (including password field, data/calendar, etc.) |
| 14 | **textarea** | multi line |
| 15 | **button** | if the button displays text on it, please additionally label the text of the button as type "label" (see below) |
| 16 | **label** | a small portion of text, typically one word or only few words, that are used together with another UI control like a radiobutton |
| 17 | **tabs** | intra-page tabs created using HTML/CSS/JS, not browser tabs, please place the rectangle around the tab handle |
| 18 | **scrollbar** | both intra-page e.g. inside textareas and the main scrollbar of the entire page if displayed |
| 19 | **pagination** | should span the entire pagination controls area, typically the next and previous buttons and page links |
| 20 | **link** | can be inside text (hyperlink), in navigation, etc. |

### 3.2.2 Participants

The total number of participants who performed the verification was 20 (10 male, 10 female), and their age ranged from 20 to 22 (mean = 21.1, SD = 0.45). They were next year's students of the Novosibirsk State Technical University crowd-intelligence lab, and none of them had participated in the aforementioned labeling. In a similar fashion, they volunteered to work in the project and provided informed consent. All the participants had normal or corrected to normal vision and reasonable experience with web UIs and IT. They did not report previous experience of working with labeling tools, and they were provided with a specially developed instruction.

*3.2.3 Procedure*

The labeled UIs were distributed among the 20 verifiers near evenly, but without random assignment. The verification process was performed independently for each UI element in each screenshot, so that the element's labeling could be identified as *correct* or *incorrect*. The reasons for a label to be marked as incorrect were described in the detailed instructions provided to the verifiers and included: too much empty space in the bounding box, cutting neighboring UI elements (except for nesting, as was described in the provided instruction presented in Figure 2), incorrect object class, etc. Also, for each labeled UI the verifying participant was asked to provide subjective assessment of the labeling completeness, i.e. if all the visible UI elements were labeled.

| A \ B | button | checkbox | radiobutton | textinput | textarea | link | selectbox | panel | list | table | pagination | tabs | scrollbar | label | paragraph | textblock | text | symbol | image | backgroundimage |
|---|---|---|---|---|---|---|---|---|---|---|---|---|---|---|---|---|---|---|---|---|
| button |  |  |  | Y | Y |  |  | Y | Y | Y |  |  |  |  | Y | Y |  |  | Y | Y |
| checkbox |  |  |  |  |  |  |  | Y | Y | Y |  |  |  |  | Y | Y |  |  | Y | Y |
| radiobutton |  |  |  |  |  |  |  | Y | Y | Y |  |  |  |  | Y | Y |  |  | Y | Y |
| textinput |  |  |  |  |  |  |  | Y | Y | Y |  |  |  |  |  |  |  |  |  | Y |
| textarea |  |  |  |  |  |  |  | Y | Y | Y |  |  |  |  |  |  |  |  |  | Y |
| link |  |  |  | Y | Y |  |  | Y | Y | Y |  |  |  |  | Y | Y |  |  | Y | Y |
| selectbox |  |  |  |  |  |  |  | Y | Y | Y |  |  |  |  |  |  |  |  |  | Y |
| panel |  |  |  |  |  |  |  | Y |  |  |  |  |  |  |  |  |  |  |  | Y |
| list |  |  |  |  |  |  |  | Y | Y | Y |  |  |  |  |  |  |  |  |  | Y |
| table |  |  |  |  |  |  |  | Y | Y | Y |  |  |  |  |  |  |  |  |  | Y |
| pagination |  |  |  |  |  |  |  | Y |  | Y |  |  |  |  |  |  |  |  | Y | Y |
| tabs |  |  |  |  |  |  |  | Y |  |  |  |  |  |  |  |  |  |  |  | Y |
| scrollbar |  |  |  |  |  |  |  | Y |  |  |  |  |  |  |  |  |  |  | Y | Y |
| label | Y |  |  |  |  |  |  | Y | Y | Y |  |  |  |  |  |  |  |  | Y | Y |
| paragraph |  |  |  | Y |  |  |  | Y |  | Y |  |  |  |  |  | Y |  |  | Y | Y |
| textblock |  |  |  | Y |  |  |  | Y |  | Y |  |  |  |  |  |  |  |  |  | Y |
| text |  |  |  | Y | Y |  |  | Y | Y | Y |  |  |  |  |  |  |  |  | Y | Y |
| symbol |  |  |  | Y | Y |  |  | Y | Y | Y |  |  |  |  |  |  |  |  | Y | Y |
| image |  |  |  |  | Y |  |  | Y | Y | Y |  |  |  |  |  |  |  |  | Y | Y |
| backgroundimage |  |  |  |  |  |  |  |  |  |  |  |  |  |  |  |  |  |  |  |  |

Figure 2: The allowed nesting (A in B) of the UI elements' classes in the labeling

The participants worked with a special web-based verification software that we created. Given a set of screenshot image files and corresponding label files in Pascal VOC XML serialization, it allowed to quickly navigate and verify the labeled UIs. Each screenshot was rendered with the labeled objects displayed as rectangular highlights of lighter color than non-labeled areas and the selected label was displayed above the bounding box and in a fixed header line above the image. The view would automatically scroll to the selected annotation to reduce verification effort from intra-page movements. To quickly verify large numbers of screenshots, keyboard shortcuts were made available for navigating the UIs and the labeled elements, as well as for setting the verification status of the selected UI element (correct or incorrect).

*3.2.4 Design*

From the binary correct/incorrect data for each UI element recorded by the verification software, we calculated the Precision$_T$ for each i-th trusted labeler, as the average precision per the $N_i$ processed screenshots:

$$Precision_{Ti} = \text{avg}_{N_i} \frac{correct}{correct+incorrect}. \quad (1)$$

The subjective completeness (SC) ranging from 0 (lowest) to 100 (highest) for each of $N_i$ labeled UIs was similarly averaged for each i-th trusted labeler:

$$SC_i = \text{avg}_{N_i} SC\ /100. \quad (2)$$

The ultimate quality index (Q) reflecting the performance of each i-th trusted labeler was then defined as follows:

$$Q_i = Precision_{Ti} * SC_i. \quad (3)$$

The quality index thus incorporated both precision and completeness of the labeling and was subsequently used to order the labelers in the trusted set.

### 3.3 The Crowdsourced Labeling

The objective of this session was to collect the data from crowdworkers, for the subsequent comparison with the trusted set. In order to utilize the distributional ground truth method, we needed enough UIs and UI elements to form distributions of the results for each crowdworker. So, our HIT (Human Intelligence Task) in AMT was designed accordingly, as described below.

*3.3.1 Material*

The material for the crowdsourced labeling was the same screenshots uploaded to AMT (only the PNG files). From the 497 initial screenshots, 2 were excluded due to the aforementioned technical problems.

The budget allocated for the AMT experimental session was 300 USD, in accordance with our estimation of an average UI labeling task difficulty and the required work effort of 5 minutes.

*3.3.2 Procedure (HIT)*

The labeling HIT was designed using the Crowd HTML elements provided by MTurk, based on the crowd-form and crowd-bounding-box widgets, with the screenshot URL as input parameter. The AMT crowd-bounding-box widget renders the screenshot, allows to zoom and pan it, and to create bounding boxes of the given types with keyboard shortcuts available for fast labeling of larger number of objects. HITs could be previewed and skipped by the crowd workers.

To explain the correct labeling of UI objects to the crowd workers, they were asked to read the instructions provided with the HIT. These instructions included:

- guidelines for correctly positioning the bounding boxes;
- a definition of the UI object classes available as labels (see below);
- a table describing the allowed nesting of boxes inside other objects' boxes depending on their types;
- a list of decision helpers for corner cases (e.g. image vs. backgroundimage) based on previous labeling difficulties observed with our trusted labelers.

As the distributional ground truth method targets complex large-scale crowdsourcing tasks (high number of HITs and high number of objects per HIT), we estimated the desired number of contributions per crowdworker as at least 20. To achieve this objective within our budget of 300 USD, we introduced a second HIT, called set HIT. This HIT was only available to crowd workers with a custom qualification, which we assigned to those workers who successfully completed at least 20 of our labeling HITs. The overall fair reward per screenshot R = 0.5 USD was calculated based on a minimum hourly wage of 6 USD and an average labeling time of 5 minutes. A worker completing 20 screenshots would thus receive $R_{20}$ = 10 USD. To incentivize workers to complete at least 20 labeling HITs, this reward was distributed between the labeling and the set HIT with a ratio of 1 to 10, resulting in $R_1$ = 0.05 USD reward for one labeling HIT and $R_S$ = 9 USD for the set HIT. The set HIT comprised a small questionnaire asking about the workers' age, gender, AMT and labeling experience and labeling difficulty and allowing to provide comments:

*"Please state how much you agree with the following statements. The scale from 1 to 100 represents "strongly disagree" at 1 to "strongly agree" at 100.*

- *I am experienced with Amazon Mechanical Turk*
- *I am experienced with Image Labeling*
- *It was difficult to label the UI Objects"*

Over a time span of 44 days from June 29 to August 11 2020, the labeling and set HITs were available on AMT in 4 batches of 80, 160, 160 and 97 screenshots. Within a batch, workers could submit as many labeling HITs as they wanted. To increase the diversity, however, workers who had successfully labeled 20 or more screenshots in a batch were not allowed to accept labeling HITs in the following batch.

*3.3.3 Design*

Exactly one label per bounding box and only labels from the list of pre-defined classes could be selected by the crowd workers. Similarly to the trusted labelers' session, the classes focused on the most frequent UI objects: including interactive (button, check, input, link, dropdown, navigation), non-interactive (image, backgroundimage) and container (table, panel) objects. However, the classes were re-organized and their number decreased to 10, due to the following considerations:

- generally lower motivation of the crowdworkers;
- the need to fit the list of classes above the fold in the corresponding screen area (see in Figure 3), to promote usability and prevent errors (crowdworkers overlooking some of the classes);
- to explore if the distributional ground truth method can work independently of the particular list of classes in the trusted and the testing set.

The labels and descriptions of the classes (as they were provided to the crowdworkers) are presented in Table 3.

In order to effectively use our budget and create a crowd-labeled dataset with diverse quality levels of sufficient size, the labeled screenshots submitted by the crowd workers were subject to our quick (5-10 s per screenshot) visual inspection. Using AMT's results Rejection mechanism, we would reject the contributions:

- of evidently malicious quality – i.e. empty submissions or submissions of a few non-existing objects arbitrary located across the screenshot;
- of workers misunderstanding the labeling task – e.g. only few objects, significantly less than required for complete labeling or labeling of only one object type.

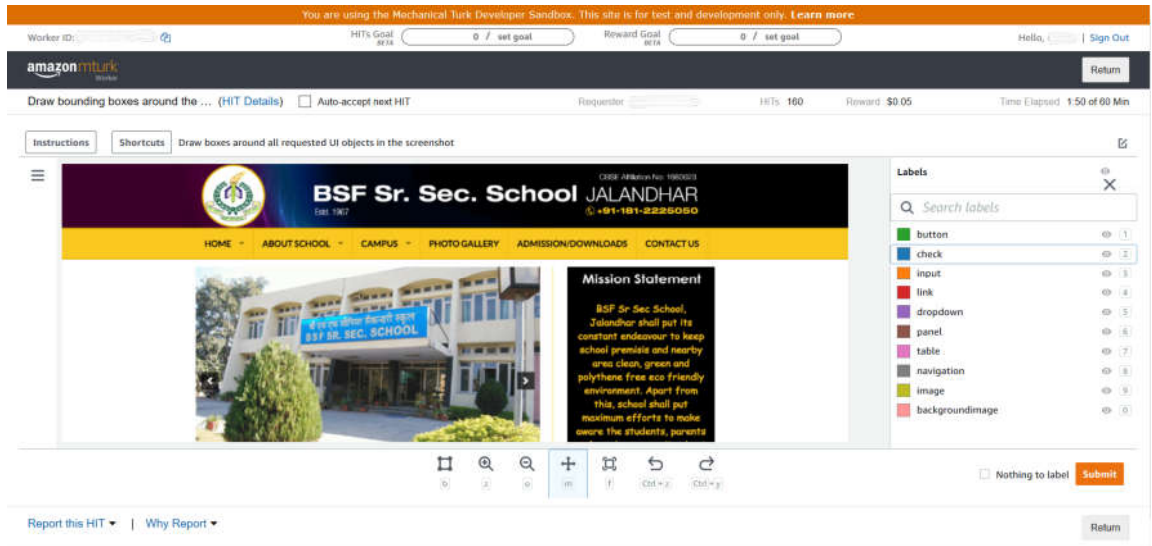

Figure 3: Screenshot of the AMT labeling interface with the list of classes

Table 3: UI elements classes used as labels in the labeling HIT

| # | Class Label | Description |
|---|---|---|
| 1 | **button** | a UI element whose visual design implies it can be **clicked** in order to **trigger an associated action** |
| 2 | **link** | a UI element whose visual design implies it can be clicked in order to **navigate** to another website/part of the website and which is **not perceived as button** |
| 3 | **check** | a square-shaped (**checkbox**) or round (**radiobutton**) UI element which can be clicked to toggle its state between **checked and unchecked** |
| 4 | **input** | **single or multi line** input for capturing alphanumerical user input, e.g. texts, numbers, passwords |
| 5 | **dropdown** | a UI element or **menu**, which would in its expanded state displays a **list of options that can be selected or multi-selected** or the visual **design** of which **implies** it **will expand** to this state when clicked |
| 6 | **table** | any visually recognizable table (using alignment, lines or background color to represent content in **rows and columns**) |
| 7 | **image** | **foreground images** that the web page displays, can **also** be smaller **symbols** like telephone icons in front of a phone number etc. |
| 8 | **backgroundimage** | images that are used as **background**, i.e. **other UI Elements** are placed **on top** of them and they have no semantic meaning |
| 9 | **navigation** | a **group** of UI element (typically links or buttons) that can be clicked to **navigate to parts of the website** (please draw one box around the entire group of UI elements that form the navigation) |
| 10 | **panel** | an **area** that is visually **separated from its surroundings** by borders, shadows, and/or background color and **contains** at least one **other UI element** |

Workers who repeatedly submitted malicious results were excluded from further submissions using the "Block Worker" mechanism. In case of a HIT rejection, an explanation was provided to the workers. All other submissions were approved and received the rewards specified above. From the UI labeling results and the data recorded by AMT, we derived the following variables for each worker:

- distribution of classes, i.e. the number of labels in each class;
- mean number of labeled elements per UI (HIT): $EUI_{AMT}$;
- mean Time-on-Task for each worker: $ToT_{AMT}$;
- $Precision_{AMT}$, as the reflection of the worker's performance, was calculated based on the number of accepted and rejected HITs for the worker, in a manner analogous to (1):

$$Precision_{AMT} = \frac{acceptedHITs}{acceptedHITs + rejectedHITs}. \quad (4)$$

*3.3.4 Participants*

The HIT was generally favorably encountered by AMT crowdworkers, with no negative comments or complaints. Some of the constructive comments collected through our questionnaire (in the set HIT) were as follows:

- *"Considering you are giving me this HIT bonus, I think you are pretty fair in validating the work done. I have experience in web design so in a way it wasn't too hard for me, but sometimes I hesitated between 2 labels (some like image links and some just plain images for example) where the answer could be very subjective, and I was wondering whether or not you would accept or reject my HIT based on the choice I picked. Thank you for this opportunity and I hope to have some more again!"*
- *"Labeling was easy as I have done this type of tasks before. The $9 target motivated me to keep on completing 20 HIT s. I was not knowing much about "Panel". … Also if possible to provide link to the website which we were working would be helpful as it was hard to guess if it was text or link as link gets Underline and it was missing in images. … Else this tool UI was great. Once we completed box it should be not movable or we should not be able to change or resize it. That will help with multiple boxes at one place. Because I worked on similar other HITs and its useful. Thank You."*
- *"More Categories should be added especially e.g. "videos" should be separated from "image" Also, I enjoy doing those HITs and hope you appreciate my accuracy of my work. If there's any future UI labelling tasks that pays more than 0.05 per HIT I'm more than happy to do it!"*

## 4 RESULTS

The dataset collected in our study is available at https://figshare.com/s/356b11d5b117014deda2. The code in Python and R is available at https://figshare.com/s/582f5aa0b564e0512f41.

### 4.1 Descriptive Statistics

*4.1.1 The Trusted Set*

In total the labelers processed 495 UIs (2 screenshots had technical problems and were removed). Their work resulted in 42716 labeled UI elements, of which 39803 (93.2%) belonged to the 20 pre-defined classes (see in Table 1). We did some minor adjustments in the erroneous custom classes (e.g., joining *textt* with *text* and *link'* with *link*). Table 4 shows the breakdown of the UI elements belonging to the pre-defined classes by the labelers (whose names are abbreviated in the column headings). As one can note from the table, the frequencies of each particular class varied between the labelers, and neither class was the most frequent in all 11 individual distributions. Therefore, outlier analysis based directly on the frequencies would not be effective.

Table 4: The distributions of the labeling classes for each trusted labeler

| Labeler / Class | AA | GD | KK | MA | NE | PV | PE | SV | SMr | SMl | VY | Total |
|---|---|---|---|---|---|---|---|---|---|---|---|---|
| image | 571 | 458 | 204 | 293 | 323 | 579 | 326 | 368 | 71 | 344 | 509 | **4046** |
| backgroundimage | 218 | 36 | 78 | 34 | 30 | 48 | 23 | 71 | 50 | 62 | 23 | **673** |
| panel | 34 | 15 | 27 | 221 | 99 | 19 | 11 | 22 | 1 | | 124 | **573** |
| list | 15 | 42 | 39 | 37 | 26 | 10 | 12 | 18 | | 106 | | **305** |
| table | 15 | 1 | | 2 | | 3 | | 2 | | 1 | | **24** |
| paragraph | 164 | 280 | 148 | 312 | 98 | 409 | 258 | 128 | 11 | 23 | 133 | **1964** |
| textblock | 65 | 9 | 13 | 42 | 6 | 37 | 48 | 239 | 3 | 260 | 6 | **728** |
| text | 1073 | 321 | 931 | 619 | 708 | 945 | 623 | 375 | 612 | 720 | 1022 | **7949** |
| symbol | 87 | 5 | 302 | 659 | 97 | | 96 | 5 | 82 | 462 | 8 | **1803** |
| checkbox | 3 | | | | 1 | 2 | | 4 | | | | **10** |
| radiobutton | 1 | 3 | 116 | | 52 | | | 20 | 4 | 1 | 2 | **199** |
| selectbox | 44 | 80 | 96 | 83 | 236 | 76 | 80 | 63 | | 47 | 58 | **863** |
| textinput | 51 | 36 | 38 | 27 | 36 | 41 | 28 | 43 | 40 | 33 | 2 | **375** |
| textarea | 14 | | | 9 | 14 | 7 | | 1 | 1 | 1 | 22 | **69** |
| button | 661 | 55 | 253 | 286 | 177 | 225 | 151 | 280 | 204 | 198 | 81 | **2571** |
| label | 253 | 368 | 152 | 632 | 948 | 181 | 211 | 226 | 181 | 8 | 128 | **3288** |
| tabs | 237 | 25 | 22 | 36 | | 39 | 25 | 20 | 3 | 10 | 10 | **427** |
| scrollbar | 10 | | | 3 | | 1 | 3 | 19 | | | 23 | **59** |
| pagination | 5 | | | 12 | | 1 | | 19 | 14 | | 41 | **92** |
| link | 1030 | 1564 | 1211 | 2042 | 1680 | 1670 | 608 | 1322 | 405 | 990 | 1263 | **13785** |
| UIs labeled | 56 | 44 | 44 | 44 | 44 | 44 | 43 | 44 | 45 | 43 | 44 | **495** |
| UI elements labeled (incl. custom classes) | 4896 | 3520 | 3927 | 5349 | 4994 | 4659 | 2649 | 3929 | 1781 | 3266 | 3746 | **42716** |
| $EUI_T$ | 87.4 | 80.0 | 89.3 | 121.6 | 113.5 | 105.9 | 61.6 | 89.3 | 39.6 | 76.0 | 85.1 | **86.3** |

During the verification of the labeling, two more UIs (0.4%) were removed from further analysis due to technical problems with the XML files, so 493 UIs remained in the analysis (the data for the affected labelers were updated accordingly). For them the 20 verifiers provided 37574 correct and 4977 incorrect ratings for the labeled UI elements, as well as the SC assessments for 487 UIs (for another 6, SC wasn't specified). In Table 5 we show the statistics for the trusted labelers ordered by the quality index (3) that incorporates both $Precision_T$ and SC. Their order in the trusted set will be considered in the subsequent DGT method application.

Pearson correlation between $Precision_T$ and SC per labelers was barely significant: $r_{11} = 0.608$, $p = 0.047$. It suggests these two components in UI labeling are interlinked, but still rather distinct. The correlation between $EUI_T$ and SC was not significant ($r_{11} = 0.496$, $p = 0.121$), which might suggest that the "true" number of elements in UIs is variable, even after averaging in reasonably large samples (43-55 UIs). Neither was $EUI_T$ significantly correlated with $Precision_T$ ($r_{11} = 0.478$, $p = 0.137$) or Q ($r_{11} = 0.461$, $p = 0.154$).

### 4.1.2 The AMT Set

Altogether 298 recorded workers participated in the 4 labeling batches (20 of them we had to block as malicious). According to the geo information provided by AMT, ¾ of the workers came from the 3 countries: US (44.8%), Brazil (15.5%) and India (13.8%); see Figure 4 for more detail.

Table 5: The results of the trusted labelers' quality verification

| # in the trusted set | Labeler | Verified UIs | Precision$_T$ | | SC | | Quality index (Q) |
|---|---|---|---|---|---|---|---|
| | | | Mean | SD | Mean | SD | |
| 1 | VY | 43 | 0.928 | 0.150 | 95.5 | 7.0 | 0.886 |
| 2 | SV | 44 | 0.974 | 0.056 | 80.4 | 12.9 | 0.783 |
| 3 | KK | 44 | 0.944 | 0.105 | 82.5 | 11.5 | 0.779 |
| 4 | GD | 44 | 0.899 | 0.078 | 84.3 | 8.1 | 0.758 |
| 5 | PV | 44 | 0.916 | 0.180 | 81.7 | 17.1 | 0.748 |
| 6 | SMl | 43 | 0.895 | 0.078 | 77.5 | 11.6 | 0.694 |
| 7 | NE | 44 | 0.851 | 0.197 | 78.3 | 24.2 | 0.666 |
| 8 | AA | 55 | 0.890 | 0.151 | 73.0 | 15.1 | 0.649 |
| 9 | PE | 43 | 0.779 | 0.147 | 72.0 | 17.2 | 0.561 |
| 10 | MA | 44 | 0.720 | 0.136 | 75.1 | 12.2 | 0.541 |
| 11 | SMr | 45 | 0.959 | 0.082 | 56.0 | 29.0 | 0.537 |
| **Total/mean:** | | **493** | **0.887** | **0.149** | **77.7** | **18.7** | **0.698** |

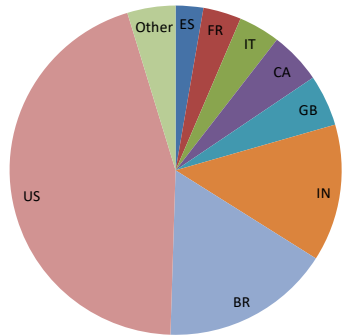

Figure 4: Geographic breakdown of the crowdworkers in our AMT session

In total, we collected 31676 labeled UI elements for 488 accepted and 754 rejected HITs. The rejection reasons were as follows (one reject can combine several reasons, so they do not sum up to 754):

- 'incomplete labeling - there are significantly more objects in the screenshot': 415;
- 'groups of objects labeled together instead of individually': 239;
- 'imprecise bounding boxes': 207;
- 'randomly labeled non-existing objects': 178;
- 'empty submission': 139;
- 'wrong object types labeled': 79.

In Figure 5 we show the relative frequencies of the 10 classes in the accepted and the rejected HITs (the classes in the horizontal axis are ordered by their total frequency in the AMT set). One can note that the two distributions differ considerably – particularly, the malicious UI labelings have comparatively more elements of infrequent classes, such as *backgroundimage* or *check*.

The mean Precision$_{AMT}$ was 0.442 (SD = 0.475), as opposed to the mean Precision$_T$ = 0.887 (SD = 0.149) in the trusted set (even though the verification process in the latter was considerably more thorough), which reinforces the need for the crowdsourcing data quality control. The mean $EUI_{AMT}$ was 28.2 (SD = 21.1), i.e. 3 times lower than $EUI_T$. The mean number of UI elements in accepted HITs was 58.3, still 1.6 times lower than the respective number in the trusted set. The correlation between the $EUI_{AMT}$ and $Precision_{AMT}$ per workers turned out to be highly significant ($r_{298} = 0.751$, $p < 0.001$), unlike in the trusted set ($r_{11} = 0.478$).

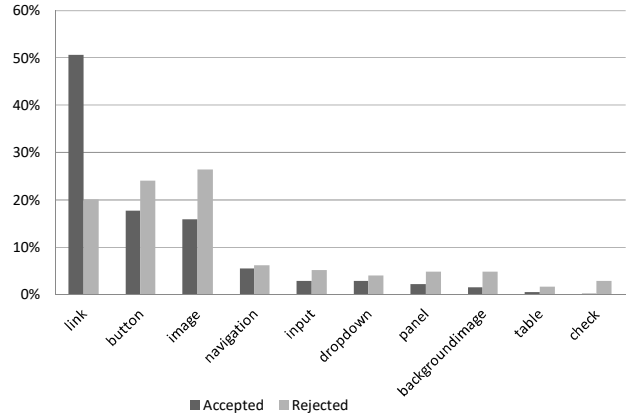

Figure 5: The classes' distributions in the accepted and rejected UI labelings

The total amount of time spent on the 1242 HITs by all the workers was 665322 s, and on average a worker devoted 635 s (SD = 481 s) to a UI labeling HIT, the correlation between $ToT_{AMT}$ and $Precision_{AMT}$ being significant ($r_{298} = 0.449$, $p < 0.001$), but considerably lower than for $EUI_{AMT}$. The correlation between $ToT_{AMT}$ and $EUI_{AMT}$ was also significant, but not as high as one might expect ($r_{298} = 0.580$, $p < 0.001$). The mean Time-on-Task turned out to be more than twice as long compared to the 5 minutes (300 s) that we estimated when planning the crowdsourcing session budget. Interestingly, 22 workers who didn't label a single UI element still spent 188416 s on the HITs, which might suggest that the Time-on-Task became widely known as a quality control parameter in crowdsourcing and can be manipulated by malicious workers.

On average, a worker attempted to perform 4.17 HITs (SD = 4.50), of which 1.64 (SD = 2.13) would be accepted. Contrary to our expectations, the extra 9 USD award for labeling 20 UIs seemingly did not motivate the workers enough, as only 9 of them have reached this threshold. Our total expended budget for the session was 126.72 USD, which corresponds to 0.26 USD per accepted labeled UI or about 0.69 USD per working (or slacking) hour.

*4.1.3 The Testing Sets*

Since the number of crowdworkers who tried to complete at least 20 HITs turned out to be lower than expected, we decided to soften the requirements for the inclusion to the DGT method testing set. The rule we applied was that a worker must have attempted at least 10 HITs (accepted or rejected) **and** have labeled at least 100 UI elements, so that a reasonably representative distribution of classes could be composed.

Of all the recorded workers, only 20 (6.71%) have complied with the rule, but it was them who provided 272 (55.7%) of all accepted UIs and 17067 (53.9%) of all labeled elements, spending 169768 s (25.5%) and earning 94.7 USD (74.7%) in total. According to the geo data provided by AMT, of these workers 35% were from the US, 30% from Brazil and the others from France (10%), Great Britain (10%), India (10%) and Canada (5%). According to the information that 9 of them provided in the set HIT questionnaire, they were male (100%) and aged 20-43 (mean = 25.1, SD = 3.98). The mean ratings on the 0 to 100 scales were:

- of their general experience with AMT: 54.8 (SD = 27.1), being close to the scale average;
- of their experience with image labeling: 63.8 (SD = 22.5), which suggest reasonably experienced workers pursuing our HITs;
- of the perceived mean difficulty of the proposed UI labeling task: 24.9 (SD = 14.3), which seems rather unchallenging.

In the subsequent sub-chapter which is dedicated to the investigation of our DGT method's effectiveness in predicting performance in crowdworkers, we are exploring different sizes of the trusted set, ranging from 1 to 9. The trusted labelers are included to the trusted set of a particular size in the order defined by their quality index (see in Table 5): e.g. {VY, SV, KK} for size 3. The screenshots labeled by the included trusted labelers **are removed from the testing set**, while the remaining ones form the testing sub-set, for which the distribution and the workers' performance are re-calculated. In other words, **there is never redundancy**: in each setup, the sets of screenshots processed by the trusted labelers and the crowdworkers do not overlap. In Table 6, we show the descriptive statistics of the testing (sub-)sets (the number of labelers = 0 corresponds to the full testing set, which is included for reference only).

Table 6: The descriptive statistics of the testing sub-sets used in the DGT investigation

| | Trusted set | | | Testing (sub-)set | | |
|---|---|---|---|---|---|---|
| # of labelers | # of UIs | % of UIs | # of workers | accepted HITs | rejected HITs | Precision (SD) |
| 0 | - | - | 20 | 272 | 205 | 0.566 (0.453) |
| 1 | 43 | 8.79% | 19 | 253 | 175 | 0.595 (0.439) |
| 2 | 87 | 17.72% | 18 | 223 | 159 | 0.584 (0.454) |
| 3 | 131 | 26.68% | 17 | 201 | 127 | 0.621 (0.438) |
| 4 | 175 | 35.64% | 14 | 167 | 108 | 0.611 (0.436) |
| 5 | 219 | 44.60% | 12 | 145 | 72 | 0.709 (0.355) |
| 6 | 262 | 53.25% | 9 | 92 | 61 | 0.612 (0.408) |
| 7 | 306 | 62.20% | 5 | 71 | 8 | 0.900 (0.120) |
| 8 | 360 | 74.53% | 4 | 42 | 7 | 0.855 (0.145) |
| 9 | 403 | 81.91% | 2 | 22 | 0 | 1.000 (0.000) |

### 4.2 The Distributional Ground Truth Method

Our DGT method relies on the assumption that the distribution of classes in UI labeling tasks is indicative of the overall performance, operationalized as $Precision_{AMT}$ in our study. As we demonstrate in Table 7, neither class was the most popular for all the workers, just as for the trusted labelers (see Table 4). In Figure 6 we show the distributions of the classes for the 20 workers of the tested set and the overlaid distribution of the best trusted labeler (VY), which resemble power law distributions. For each of the workers, the classes were sorted by frequency, so the ranks on the horizontal axis may correspond to different classes. Then they were normalized by dividing each value by the mean frequency for the worker. For the trusted labeler, the same procedure was performed, and also every two frequencies were averaged to one rank and value in the diagram (as trusted labelers had 20 pre-defined classes instead of 10 classes for the workers).

The actual DGT method is straightforward now. For each tested worker, we take the distribution of the classes and apply the two-sample Kolmogorov-Smirnov test that compares it to the distribution of each labeler in the trusted set. The p-values produced by the tests are averaged for each worker and are used to predict precisions in the testing sub-set.

Table 7: The aggregated distribution of the labeling classes in the testing set (20 workers) and the full AMT set

| Class | Testing set | | AMT set |
|---|---|---|---|
| | overall frequency | # times most frequent | overall frequency |
| link | 8604 | 12 | 14524 |
| button | 3134 | 2 | 5933 |
| image | 3036 | 6 | 5554 |
| navigation | 885 | | 1773 |
| panel | 362 | | 838 |
| dropdown | 359 | | 943 |
| input | 330 | | 1023 |
| backgroundimage | 244 | | 659 |
| table | 86 | | 239 |
| check | 27 | | 190 |
| **Total** | **17067** | **20** | **31676** |

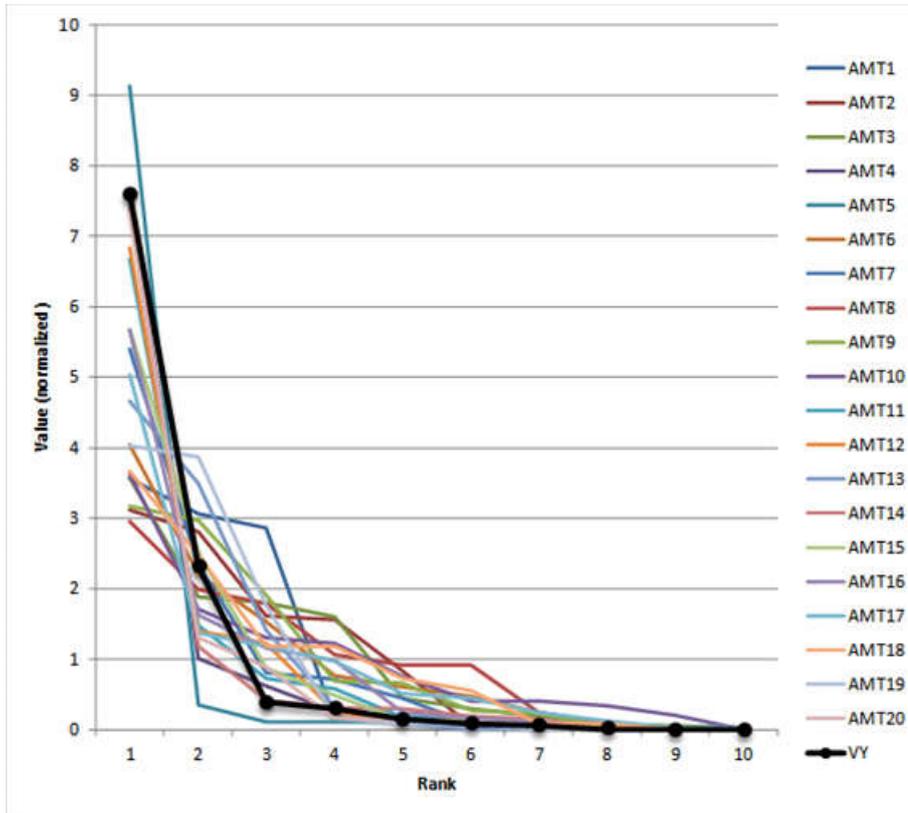

Figure 6: The distributions of labeling classes in the selected workers (AMT$_i$) and the best trusted labeler (VY)

So, we applied the method for all the sub-sets previously specified in Table 6 (even though with low number of workers in the testing sub-set the model makes little sense) and present the outcomes in Table 8. The best $R^2 = 0.855$ corresponding to the trusted set size = 2 (trusted labelers VY and SV) is highlighted.

Table 8: The crowdworkers' performance prediction models resulting from the DGT method

| Size of trusted set | Size of testing sub-set | The model | |
|---|---|---|---|
| | | $R^2$ | F, p |
| 8.79% (1 trusted labeler) | 91.21% (19 workers) | 0.658 | $F_{1,17}$ = 32.7, p < 0.001 |
| 17.72% (2) | 82.28% (18) | **0.855** | $F_{1,16}$ = 94.5, p < 0.001 |
| 26.68% (3) | 73.32% (17) | 0.789 | $F_{1,15}$ = 56.0, p < 0.001 |
| 35.64% (4) | 64.36% (14) | 0.716 | $F_{1,12}$ = 30.3, p < 0.001 |
| 44.60% (5) | 55.40% (12) | 0.539 | $F_{1,10}$ = 11.7, p = 0.007 |
| 53.25% (6) | 46.75% (9) | 0.789 | $F_{1,7}$ = 26.1, p = 0.001 |
| 62.20% (7) | 37.80% (5) | 0.107 | $F_{1,3}$ = 0.4, p = 0.591 |
| 74.53% (8) | 25.47% (4) | 0.501 | $F_{1,2}$ = 2.0, p = 0.292 |
| 81.91% (9) | 18.09% (2) | - | - |

### 4.3 Baseline and Alternative Models

To evaluate our DGT model (trusted set size = 2), we compared it to the baseline and considered some other alternative factors (see in Table 9). Particularly, we calculate $R^2$s in regressions built for $Precision_{AMT}$ for the following variables:

- $ToT_{AMT}$ – the baseline, often used in crowdsourcing quality control;
- attempted HITs (accepted + rejected) – the factor that nominally reflects the involvement of a worker in our study;
- $EUI_{AMT}$ – the factor that arguably best reflects the actual work effort by a worker in our UI labeling HITs;
- $GOF_{PL}$ – the goodness-of-fit measure of a worker's classes distribution to the power law distribution (obtained using the third-party plpva.r library (http://tuvalu.santafe.edu/~aaronc/powerlaws/plpva.r) that implements the test described in [26].

Table 9: The detailed results for the predictive model for $Precision_{AMT}$ (trusted set size = 2)

| Worker's # in the sub-set | $Precision_{AMT}$ | p-values from the KS test: | | | Alternative factors | | | |
|---|---|---|---|---|---|---|---|---|
| | | with VY | with SV | Avg. | attempted HITs | $ToT_{AMT}$ | $EUI_{AMT}$ | $GOF_{PL}$ |
| 1 | 0.974 | 0.856 | 0.837 | 0.847 | 39 | 191 | 56.31 | 0.492 |
| 2 | 0 | 0.091 | 0.158 | 0.124 | 34 | 50 | 4.62 | 0.617 |
| 3 | 1 | 0.276 | 0.937 | 0.606 | 32 | 558 | 70.13 | 0.706 |
| 4 | 0.813 | 0.686 | 0.987 | 0.837 | 32 | 325 | 37.16 | 0.640 |
| 5 | 0.731 | 0.974 | 0.704 | 0.839 | 26 | 102 | 24.00 | 0.589 |
| 6 | 0 | 0.002 | 0.002 | 0.002 | 25 | 63 | 5.24 | 0.115 |
| 7 | 0 | 0.066 | 0.012 | 0.039 | 23 | 126 | 4.39 | 0.354 |
| 8 | 0 | 0.458 | 0.158 | 0.308 | 19 | 77 | 9.21 | 0.562 |
| 9 | 0 | 0.019 | 0.023 | 0.021 | 19 | 94 | 6.11 | 0.406 |
| 10 | 1 | 0.482 | 0.517 | 0.499 | 18 | 619 | 57.72 | 0.640 |
| 11 | 1 | 0.686 | 0.704 | 0.695 | 18 | 232 | 39.06 | 0.600 |
| 12 | 1 | 0.608 | 0.875 | 0.741 | 16 | 1370 | 60.81 | 0.592 |
| 13 | 1 | 0.987 | 0.837 | 0.912 | 16 | 568 | 65.19 | 0.529 |
| 14 | 1 | 0.913 | 0.751 | 0.832 | 14 | 427 | 71.43 | 0.627 |
| 15 | 1 | 0.738 | 0.837 | 0.788 | 14 | 1326 | 68.43 | 0.639 |
| 16 | 0 | 0.259 | 0.032 | 0.146 | 14 | 57 | 7.21 | 0.296 |
| 17 | 1 | 0.913 | 0.751 | 0.832 | 12 | 837 | 76.83 | 0.659 |
| 18 | 0 | 0.003 | 0.010 | 0.006 | 11 | 355 | 9.27 | 0.431 |
| $R^2$ for $Precision_{AMT}$: | | 0.658 | 0.895 | 0.855 | < 0.01 | 0.401 | 0.875 | 0.480 |

The results presented in the table suggest that $R^2 = 0.875$ ($F_{1,16} = 112.0$, $p < 0.001$) for $EUI_{AMT}$ was marginally higher than $R^2 = 0.855$ ($F_{1,16} = 94.5$, $p < 0.001$) for our DGT model, although somehow lower than $R^2 = 0.895$ ($F_{1,16} = 135.8$, $p < 0.001$) for one of the trusted labelers (SV) in the model. The $GOF_{PL}$ factor was considerably less compelling ($R^2 = 0.480$, $F_{1,16} = 14.8$, $p = 0.001$), but still superior to the baseline $ToT_{AMT}$ ($R^2 = 0.401$, $F_{1,16} = 10.7$, $p = 0.005$). The regression model for $Precision_{AMT}$ with both factors, the average p-value (Beta = 0.472, $p = 0.001$) and the $EUI_{AMT}$ (Beta = 0.539, $p < 0.001$), had further enhanced $R^2 = 0.941$ ($F_{2,15} = 118.8$, $p < 0.001$). We are going to appraise this in the Discussion section.

To verify the effectiveness of the DGT method with our initially planned minimal number of 20 attempted HITs per worker, we sampled the 7 workers that comply with this rule from the testing sub-set corresponding to the trusted set size = 2. In this sample, $R^2 = 0.869$ for our model somehow increased, but $R^2 = 0.860$ for the $EUI_{AMT}$ factor diminished. We have also calculated the respective values for the entire AMT set. For $EUI_{AMT}$, the $R^2 = 0.564$ ($F_{1,296} = 383.1$, $p < 0.001$) was notably higher than $R^2 = 0.202$ ($F_{1,296} = 74.9$, $p < 0.001$) for $ToT_{AMT}$. However, for the trusted set, the $EUI_T$ factor was not predictive of either $Precision_T$ ($R^2 = 0.228$, $F_{1,9} = 2.66$, $p = 0.137$) or Q ($R^2 = 0.212$, $F_{1,9} = 2.43$, $p = 0.154$).

## 5 DISCUSSION AND CONCLUSIONS

The results of the DGT method exploration imply that it might be applicable for the crowdsourcing data quality control in the UI labeling tasks that we considered. The $R^2$s produced by the method were of 0.8 and higher for the reasonably practical ratios between the trusted and the testing sets' sizes, 17-27% (Table 8). The ability of the method to predict performance in crowdworkers was considerably higher than that of the Time-on-Task factor ($R^2 = 0.401$) that is traditionally used for this purpose and that we considered as a baseline. The latter was surpassed even by the $GOF_{PL}$ factor ($R^2 = 0.480$) based on comparison of the workers' distributions of classes to power law distribution. It actually might provide an interesting alternative to the DGT method, as $GOF_{PL}$ requires no trusted set at all, but this should be the objective of another research work with larger distributions of classes.

Meanwhile, an alternative factor $EUI_{AMT}$ that we considered showed somehow superior $R^2$ compared to the DGT model's $R^2$s in some of the testing setups (Table 9). This is understandable, since $EUI_{AMT}$ in our HIT was the best reflection of the work effort contributed to the task. We would however argue that the number of elements per UI is easily prone to malicious manipulations, similarly to the once indicative Time-on-Task. The latter in our study was inflated even by the workers who did not label a single UI element thus was not performing an actual task. Similarly, increased $EUI_{AMT}$ could be futilely exaggerated with relatively little effort, e.g. through random specification of labels, possibly even with browser automation scripts. On the contrary, we see **no uncomplicated way to imitate a trustworthy distribution of the classes**, since malicious workers would have no idea about the characteristics of the trusted set. Also, in the trusted set that corresponds to higher-quality labeling data, the effect of $EUI_T$ on either $Precision_T$ or Q was not significant, which questions the true impact of the factor. Finally, in the regression models with the two factors, both of them were significant and had comparable Betas, so we should assume that the classes' distributions and the mean number of elements per UI reflect two different facets of the crowdsourced UI labeling.

Another issue worth discussing is whether the effectiveness of the DGT method is due to the KS test considering mostly the **locations** of the distributions in our context. Indeed, the mean $EUI_T = 86.3$ was a great deal higher than the mean $EUI_{AMT} = 28.2$, and the mean number of UI elements in accepted HITs (58.3) would be closer to $EUI_T$. We however argue that the effect of precision in trusted labelers with respect to explaining the workers' $Precision_{AMT}$ was more prominent than the effect of completeness as expressed by $SC_i$. Indeed, of the considered two trusted labelers (see in Table 9), SV had higher $R^2 = 0.895$ than VY's $R^2 = 0.658$, notably lower $SC_{SV} = 80.4$ in comparison to $SC_{VY} = 95.5$, but higher Precision$_{SV}$ = 0.974 vs. Precision$_{VY}$ = 0.928. It might suggest that the effectiveness of the DGT method was mostly due to the distributions' **shapes**, though surely this statistically unrepresentative example calls for further investigation.

A related and very practical issue is the desired number and quality of trusted labelers. Our assumption that trusted labelers with greater quality index would have "better" distributions was not confirmed in practice, as demonstrated by VY's and SV's $R^2$'s in Table 9. So, we plan to explore efficient approaches for composing trusted sets for the DGT method in our further research work. Currently, we would just recommend having reasonable diversity of trusted labelers and assume that the averaged p-values negate the effect of individual discrepancies.

Also, we must recognize that at the current stage of our research we are unable to quantitatively express, beyond an estimation, the advantage of the DGT method over the alternatives based on redundancy, such as the widely used GT and MC. In our context, one would probably have a GT of size 1, i.e. one completely and correctly labeled UI screenshot, and make sure that every worker has to label this, in a sort of an entrance-test to other HITs. Provided that in our AMT experimental session a worker on averaged labeled 4.17 screenshots, this would correspond to the **quality control process leading to wasting 24% of the outcome**. In case of MC implying that at least 3 workers label a screenshot, the share of the largely unused results would be even greater, at 67%.

In turn, the DGT method has certain limitations. Arguably the strongest one is that **a worker needs to produce enough results to compose a representative distribution of the classes** – in our study, at least 100 UI elements labeled in 10 UIs. Indeed, the excluded workers contributed 216 (44.3%) of accepted HITs, which could not be covered by the method and would probably need to undergo different quality control procedures. However, one should consider that our experiment was artificially set up with a limited number of screenshots, whereas in real circumstances HIT design would be different. Moreover, the UI labeling task has an entry threshold – the workers need to comprehend the classes, read instructions, etc., so the learning effect is a positive thing and fewer workers each performing more HITs should be preferred to the contrary situation. Another limitation is that the trusted set might be bound to UIs belonging to a particular domain, and the transferability of the trusted distributions to other domains (e.g. from websites of universities to museums) is so far unexplored.

Finally, we need to note that the assumptions for the two-sample KS test were not totally satisfied in our study. The variables (classes' distributions) were not continuous and the number of their values was rather modest (although ranging in a large interval). E.g. in [27] it is noted that for small sample sizes the nominal significance of 0.1 corresponds to the actual significance of 0.0835. However, the effects that we found in our study were rather strong and we probably can assume the findings are statistically valid. Also, we did try the *ks.boot* function in R that is considered an alternative to *ks.test* (adding simulation), but it did not produce any different p-values for our data.

So, the contributions of our work can be summarized as follows:

- we proposed the Distributional Ground Truth method for crowdsourcing data quality control, which implies zero redundancy, thus having the potential to obviate the ancillary work effort and expenses;
- we demonstrated that shapes of classes' distributions (labels' frequencies) are reflective of the overall crowdworkers' performance in UI labeling tasks;
- we found significant relation between the distributions' goodness-of-fit to power law and the UI labeling precision, which might hint on the currently underexplored similarity between real web UIs and non-random natural language texts.

Our further research prospects include exploration of the method's applicability: whether it could be feasible in other crowdsourcing tasks, what are the efficient approaches for composing the trusted set, etc. However, even at the current stage of development we hope that our results can contribute to more efficient non-redundant crowd data quality control and thus to better utilization of human mind power in HCI-related ML tasks.